\documentclass{svjour3}                     
\smartqed  
\usepackage{graphicx}
\usepackage{fix-cm}

\journalname{Journal of Low Temperature Physics}
\begin{document}

\title{On role of symmetries in Kelvin wave turbulence
}


\author{V. V. Lebedev \and  V. S. L'vov \and   S. V.~Nazarenko
}


\institute{Vladimir V. Lebedev \at
              Landau Institute for theoretical physics RAS,
 Moscow, Kosygina 2, 119334, Russia \\
              \email{lebede@itp.ac.ru}   \\
            Victor S. L'vov  \at
            Department of Chemical Physics, The Weizmann Institute of Science, Rehovot 76100, Israel\\
            \email{victor.lvov@gmail.com}  \\
           Sergei Nazarenko
            \at Warwick
}

\date{Received: date / Accepted: date}

\maketitle

The argumentation of  E.V. Kozik and B.V. Svistunov (KS) in their Comment~\cite{KS-10c}
on Notes ``\emph{Symmetries and Interaction coefficients of Kelvin waves}" by two of us
(VVL \& VSL)~\cite{LL-10}   relies mainly on the KS text ``\emph{Geometric Symmetries in
Superfluid Vortex Dynamics}''~\cite{KS-10}. Therefore we start with commenting this text.

First part of Ref.~\cite{KS-10} reproduces the KS theoretical scheme of the Kelvin waves
cascade investigation presented, e.g., in Refs.~\cite{KS-04,KS-rev}. Next, they discuss
separation of fast and slow variables introducing a fast field $w$ in the local reference
system attached to the vortex position determined by the slow variable. Then, in complete
accordance with our derivation, presented earlier in~\cite{LL-10}, KS concluded that
vertices describing interaction of slow and fast variables depend on the vortex curvature
determined by the slow variable, i.e. that the vertices are proportional to the second
power of the small wave vector. This is treated by KS as a proof of the cascade locality.
However, as we stressed in~\cite{LL-10}, the interaction vertices in the kinetic equation
for the Kelvin waves are not the same as obtained for the above variable $w$. The
vertices in the kinetic equation are derived from a direct expansion of the Hamiltonian
in a series over the action variables (normal coordinates). And, as we demonstrated in
Ref.~\cite{LL-10}, the symmetry does not forbid linear in $k$ asymptotic behavior of such
vertices.

Now some remarks concerning the specific KS assertions, made in~\cite{KS-10c}:
\begin{description}

\item ``In fact, the naiveness of the authors of Ref.~\cite{LL-10} is in failing
to appreciate that, being intermediate auxiliary concepts, the bare vertices themselves
have no direct physical meaning and thus are not supposed to respect the tilt invariance
on the individual basis.''

This is a strange conclusion since the symmetry arguments are correct for any quantity
possessing corresponding (in our case  rotational) symmetry. Therefore they are equally
applied to the Hamiltonian and to the normal symbol of the $S$-matrix (expansion of which
generates the complete interaction vertices). For illustration, we have chosen the vortex
length $L$ as the simplest rotationally invariant object. Its properties, that can be
easily established by the direct calculations, represent general features of the
rotationally invariant quantities.\vskip .2cm 

\item ``Analogously, in a more general case of the full Biot-Savart model, the invariance of
dynamics with respect to the shift and tilt of the vortex line prescribes that, after
combining the corresponding vertices $T^{3,4}_{1,2}$ and $W^{4,5,6}_{1,2,3}$ into $V^
{4,5,6}_{1,2,3}$, the terms $\propto k_1k_2k_3k_4k_5k_6$ at $k_1\ll k_{2,...,6}$ must
cancel exactly leaving the first non-vanishing contribution to $V^{4,5,6}_{1,2,3}$
proportional to $k^2_1$, which was demonstrated in our Ref. [3].''

Nothing of the kind was proved by KS, neither in the present nor in the previous paper.
KS only showed absence of the linear $k$-asymptotics in the interaction vertex for the
\em local \em amplitudes $w$, and not for the \em action variables \em of the basic
Hamiltonian. And, again, their local analysis has no direct relation to the vertices in
the kinetic equation.

\end{description}

Finally, we would like to say that we are surprised by lighthearted attitude of KS to the
results of direct calculation of the interaction vertex $V^{4,5,6}_{1,2,3}$ in the full
Biot-Savart case done in Ref.~\cite{LLNR-10}. This paper has fixed serious mistakes in
the leading order of $V^{4,5,6}_{1,2,3}$ made in the previous papers of KS, and for the
first time completed the analytical calculation of this vertex. Since, as we showed in
\cite{LL-10}, the symmetry arguments do not prevent the linear $k$-asymptotic in
$V^{4,5,6}_{1,2,3}$, the direct calculation appears to be the only way to find the
leading-order asymptotic behavior in order to analyze the locality property.

\vskip 0.2 cm

\textbf{\emph{To conclude}}, there is neither proof of locality nor any refutation of the
possibility of linear asymptotic behavior of interaction vertices in the
texts~\cite{KS-10c,KS-10} of Kozik and Svistunov.

\end{document}